# Financial Analysis： Intelligent Financial Data Analysis System Based on LLM-RAG

**Jingru Wang[1,4], Wen Ding[2,5], Xiaotong Zhu[3,6]**

[1] University of Pennsylvania, Pennsylvania, USA
[2] H. Milton Stewart School of Industrial and Systems Engineering,
Georgia Institute of Technology, Atlanta, GA, USA
[3] Tepper School of Business, Carnegie Mellon University, Pittsburgh, PA, USA

[4] 1812503968@qq.com
[5] wding5662@gmail.com
[6] xiaotonz@alumni.cmu.edu

**Abstract.** In the modern financial sector, the exponential growth of data has made efficient and accurate financial data analysis increasingly crucial. Traditional methods, such as statistical analysis and rule-based systems, often struggle to process and derive meaningful insights from complex financial information effectively. These conventional approaches face inherent limitations in handling unstructured data, capturing intricate market patterns, and adapting to rapidly evolving financial contexts, resulting in reduced accuracy and delayed decision-making processes. To address these challenges, this paper presents an intelligent financial data analysis system that integrates Large Language Models (LLMs) with Retrieval-Augmented Generation (RAG) technology. Our system incorporates three key components: a specialized preprocessing module for financial data standardization, an efficient vector-based storage and retrieval system, and a RAG-enhanced query processing module. Using the NASDAQ financial fundamentals dataset from 2010 to 2023, we conducted comprehensive experiments to evaluate system performance. Results demonstrate significant improvements across multiple metrics: the fully optimized configuration (gpt-3.5-turbo-1106+RAG) achieved 78.6% accuracy and 89.2% recall, surpassing the baseline model by 23 percentage points in accuracy while reducing response time by 34.8%. The system also showed enhanced efficiency in handling complex financial queries, though with a moderate increase in memory utilization. Our findings validate the effectiveness of integrating RAG technology with LLMs for financial analysis tasks and provide valuable insights for future developments in intelligent financial data processing systems.

**Keywords:** Intelligent financial analysis; LLM-RAG; NASDAQ data; Model comparison; Technology optimization

## 1. Introduction

In the modern financial landscape, the exponential growth of data presents both opportunities and challenges for decision-makers. The demand for efficient and accurate financial data analysis has never been more crucial.. Traditional methods often struggle to process the vast amounts of complex and diverse financial information    within required timeframes. Fortunately, the emergence of artificial intelligence and machine learning technologies has paved the way for enhancing financial data analysis capabilities.

Large language models (LLMs) have shown remarkable potential in natural language processing tasks. However, when applied directly to the financial domain, LLMs face limitations, including insufficient domain-specific knowledge and challenges in handling specialized financial terminologies. Retrieval-augmented generation (RAG) technology has emerged as a powerful solution to overcome these challenges. By integrating external knowledge retrieval with LLMs, RAG enables the models to access relevant financial information swiftly, thereby improving their understanding and analytical capabilities.

This research focuses on developing an intelligent financial data analysis system based on LLM-RAG. The system aims to leverage the strengths of both technologies to deliver more accurate and insightful financial analysis. By utilizing a comprehensive dataset of NASDAQ financial fundamentals and designing a sophisticated model architecture, we strive to overcome the existing limitations in financial data analysis and contribute to the further advancement of the field.

**2. Literature Review**

In the domain of financial data analysis and intelligent decision-making, the combination of retrieval-augmented generation (RAG) technology and large language models (LLM) has emerged as a research hotspot. The following provides an overview of the research landscape and main achievements in this area based on the relevant references.

Arslan et al. conducted a comprehensive review and case study on the use of RAG-LLM for obtaining business insights **[1].** They delved deeply into how to enhance the performance of LLM in handling complex business data using RAG technology and demonstrated its potential in mining valuable business information and assisting decision-making through practical cases. This laid the foundation for the integration of theory and practice in subsequent research and revealed the feasibility and advantages of applying RAG-LLM in financial business scenarios.

Setty et al. focused on improving the retrieval of question answering models based on RAG for financial documents **[2].** Given the professionalism and complexity of financial data, traditional retrieval methods face challenges in RAG-based question answering models. Their research proposed innovative strategies to optimize the retrieval process, aiming to enhance the model's ability to accurately locate and extract relevant information from massive financial documents. By improving index construction and semantic matching algorithms, they increased the accuracy and efficiency of the question answering model's response to financial questions, effectively addressing the key issue of financial information retrieval.

Zhang et al. were dedicated to enhancing financial sentiment analysis using retrieval-augmented large language models **[3].** The sentiment in the financial market is crucial for investment decisions and market trend judgments. Their research integrated RAG technology with LLM to enhance the model's ability to understand sentiment trends in financial texts. By retrieving external knowledge to supplement the model's understanding of the financial context and sentiment words, they significantly improved the accuracy of sentiment analysis in actual financial text analysis, providing a more reliable tool for monitoring the sentiment of the financial market.

Wang et al. proposed OmniEval, which provides a comprehensive and automatic evaluation benchmark for RAG systems in the financial field **[4].** With the widespread application of RAG technology in financial data analysis, the lack of a unified and effective evaluation standard has become a limiting factor. OmniEval covers multi-dimensional indicators, comprehensively measuring the performance of RAG systems from aspects such as retrieval accuracy, relevance of generated content, and financial logic rationality. It provides a comparable evaluation framework for different research and applications, promoting the standardization and normalization of financial RAG technology.

Lim and Suh focused on the study of implementing the most optimized RAG system for financial documents using AutoRAG **[5].** In financial document processing, how to automatically configure and optimize the parameters of the RAG system is the key to improving application efficiency. Their research utilized AutoRAG technology to automatically adjust retrieval sources and model hyperparameters. Experiments demonstrated that it can effectively improve the performance of the system in financial document analysis tasks and reduce the cost of manual tuning, facilitating the efficient deployment and application of RAG technology in the financial field.

Mao et al. focused on the domain adaptation of retrieval-augmented generation in RAG-Studio **[6]**. The financial field has unique professional knowledge systems and language expressions, and the performance of models often degrades when applied across domains. RAG-Studio enabled the model to better adapt to the characteristics of the financial field through a self-alignment mechanism, optimizing the model's performance in understanding financial terms and reasoning about industry-specific logic, enhancing the pertinence and effectiveness of RAG technology in the financial field.

Yepes et al. conducted research on chunking financial reports to optimize retrieval-augmented generation, which is of great significance **[7]**. Financial reports have complex structures and voluminous contents, and reasonable chunking helps improve retrieval efficiency and information utilization. Their research designed a specific method for chunking financial reports and combined it with RAG technology to enable the model to quickly focus on relevant report chunks for knowledge retrieval and generation, enhancing the ability to process information in financial reports and providing innovative ideas and methods for financial data analysis.

In conclusion, significant progress has been made in the application of RAG-LLM technology within the financial filed. On-going in-depth exploration has addressed various aspects, including model performance optimization, sentiment analysis, establishment of evaluation standards, domain adaptation, and data processing, providing valuable theoretical support and practical insights for the development of intelligent financial data analysis systems. However, challenges such as multi-source data fusion and model adaptability in complex environments remain, requiring further research and breakthroughs.

## 3. Data Introduction

This research utilizes the NASDAQ Financial Fundamentals dataset obtained from the Kaggle platform, which comprises quarterly financial fundamental data for companies listed on the NASDAQ stock exchange. The dataset encompasses comprehensive financial metrics from major technology companies, including industry leaders such as Apple (AAPL), Microsoft (MSFT), and Google (GOOGL), providing a robust foundation for financial analysis and research.

The dataset spans from 2010 to 2023, offering a longitudinal perspective on corporate financial performance through quarterly reporting periods. The data structure incorporates essential fields including reporting periods, company identifiers, stock tickers, financial indicators, and corresponding monetary values. This comprehensive coverage enables detailed analysis of financial trends and corporate performance over time.

**Table 1.** Details of core variables in financial statements (NASDAQ)

| Variable Name | Data Type | Description | Example |
| --- | --- | --- | --- |
| period | datetime | The fiscal quarter end date for the financial reporting period | 2023/3/31 |
| company | string | The full company name of the listed entity | Apple Inc. |
| tickers | string | The stock market symbol/ticker used for trading | AAPL |
| indicator | string | The type of financial metric being reported | Revenue, Assets |
| amount | float | The monetary value of the financial metric (in USD) | 100000000 |

The financial metrics captured in the dataset encompasses crucial indicators such as total assets, revenue, net income, and operating income, providing a holistic view of corporate financial health. These indicators serve as fundamental parameters for assessing company performance, profitability, and operational efficiency across different time periods and market conditions.

To ensure research reliability and data integrity, systematic preprocessing procedures were implemented. These include the standardization of monetary values by eliminating currency symbols and thousand separators, normalization of date formats, and validation of data consistency. Missing values and anomalies were identified and resolved through rigorous data cleaning procedures, while maintaining the integrity of the original financial information.

The dataset exhibits substantial research significance through its comprehensive coverage of the technology sector, temporal relevance, and completeness of financial metrics. The inclusion of major technology companies provides representative insights into the financial dynamics of the technology industry, while the quarterly reporting frequency enables detailed temporal analysis of financial trends and patterns.

The longitudinal nature of the dataset facilitates the examination of financial performance evolution over time, enabling the identification of trends, seasonal patterns, and potential correlations between different financial metrics. This temporal depth, combined with the breadth of financial indicators, provides a solid foundation for both cross-sectional and time-series analyses of corporate financial performance.

Furthermore, the dataset's focus on NASDAQ-listed companies ensures enhances both the quality and reliability of the data , as these companies are bound by stringent reporting standards and regulations . This institutional framework enhances the credibility of the financial data and supports robust academic research on corporate financial performance and market dynamics within the technology sector.

This comprehensive dataset thus serves as a valuable resource for investigating corporate financial performance, market trends, and industry dynamics within the technology sector, supporting detailed analytical approaches and research methodologies in financial analysis and corporate performance assessment.

**4. Model design and results analysis**

*4.1. Model Architecture Dissection*

This module is the cornerstone of the entire system. It conducts multi-dimensional processing on the NASDAQ financial fundamentals dataset sourced from the Kaggle platform. In terms of monetary values, the removal of currency symbols and thousand separators standardizes the data, standardizing various currency units   and data representations and laying the foundation for subsequent numerical calculations and analyses. The normalization of date formats ensures the consistency of time series data, facilitating data mining and trend analysis along the time dimension. For missing values and outliers, a rigorous cleaning procedure is implemented . While maximizing the retention of the integrity of the original financial information, it improves data quality and minimizes the impact of data noise   on the analysis results, ensuring reliable data input for subsequent precise analysis.

The preprocessed data is converted into vector form for storage. This process utilizes advanced vector representation technologies such as word embedding models. Mapping financial text data to a low-dimensional vector space enables the data to retain semantic information while optimizing efficient storage and retrieval characteristics. By rationally designing the vector storage architecture, it can quickly respond to subsequent queries and retrieval requests. Based on the similarity measurement between vectors, it can rapidly locate financial data relevant to the query, effectively enhancing the retrieval efficiency of the system and providing strong support for the model to quickly access relevant information when facing complex financial problems.

This part is the key interaction component of the system. After the user inputs a query, the system, based on the vector storage structure and specific retrieval algorithms, quickly screens and extracts highly relevant information from massive financial data. The system   utilizes natural language processing technology to understand the user's query intention and combines the financial domain knowledge graph and semantic understanding model to accurately match the data, improving the relevance and accuracy of the retrieval results. It not only depends on the efficient construction of indexes and search algorithms but also necessitates continuous optimization of the query understanding model to adapt to the complex and specialized terminology in the financial sector, , providing an accurate data subset for subsequent analysis.

Figure 1 presents the architectural design of the proposed LLM-RAG based financial data analysis system, showcasing the integration of key components and their information flow.

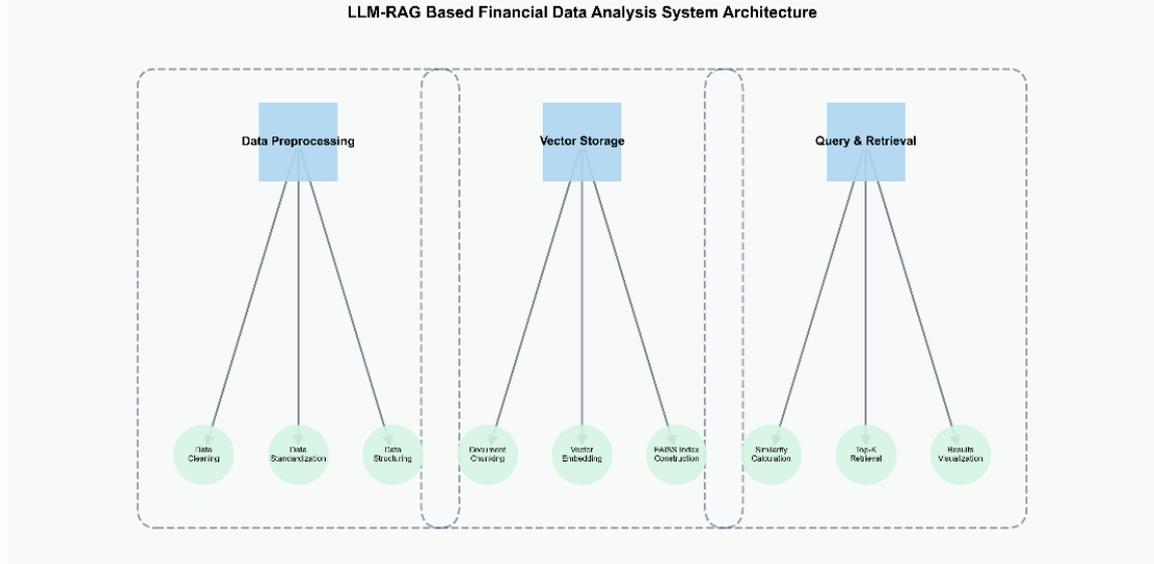

**Figure 1.** LLM-RAG Based Financial Data Analysis System Architecture

*4.2. Experimental Environment and Experimental Design*

To comprehensively evaluate the effectiveness of our proposed LLM-RAG based financial data analysis system, we conducted extensive experiments under rigorous controlled conditions. This section details our experimental methodology, evaluation metrics, and result analysis.

The experiments were conducted in a controlled environment using industry-standard computing infrastructure. The hardware configuration consisted of NVIDIA A800 GPU with 80GB HBM2e memory, complemented by 256GB system RAM, providing sufficient computational capacity for large-scale financial data processing. This setup ensures optimal performance for both model training and inference tasks while maintaining cost-effectiveness. The software environment was built on Python 3.8 with PyTorch 1.9.0 as the deep learning framework.

The experimental configuration is comprehensively documented across Tables 2-5, where Table 2 describes the hardware infrastructure including computing resources and storage specifications, Table 3 enumerates the software stack and corresponding version information, Table 4 presents the detailed model architectures and hyperparameter settings for both LLM and RAG implementations, and Table 5 outlines the data processing pipeline and associated tools utilized in our experimental framework.

**Table 2.** Experimental Environment Configuration

| Category | Specification |
|---|---|
| Operating System | Ubuntu 22.04 LTS |
| GPU | NVIDIA A800 (80GB HBM2e) |
| CPU | AMD EPYC 7763 64-Core Processor |
| | RAM- 256GB DDR4 |
| Storage | 2TB NVMe SSD |

**Table 3.** Software Environment

| Category | Specification |
|---|---|
| Python | 3.8.10 |
| PyTorch | 1.9.0 |
| CUDA | 11.8 |
| cuDNN | 8.6.0 |
| transformers | 4.30.2 |
| sentence-transformers | 2.2.2 |
| faiss-gpu | 1.7.4 |
| scikit-learn | 1.0.2 |
| pandas | 1.5.3 |
| numpy | 1.23.5 |

**Table 4.** Model Configurations

| Category | Specification |
|---|---|
| Base LLM | gpt-3.5-turbo |
| Enhanced LLM | gpt-3.5-turbo-1106 |
| Vector Embedding | all-MiniLM-L6-v2 |
| Vector Database | FAISS |
| Batch Size | 32 |
| Learning Rate | 2.00E-05 |
| Max Sequence Length | 512 |
| RAG Parameters | - |
| Retrieval Top-k | 5 |
| Context Window | 1024 |
| Vector Dimension | 384 |

**Table 5.** Data Processing

| Category | Specification |
|---|---|
| Dataset Format | JSON |
| Data Preprocessing | pandas, numpy |
| Text Tokenization | transformers.AutoTokenizer |
| Vector Indexing | FAISS.IndexFlatIP |

We implemented a systematic experimental design with four distinct groups to evaluate different aspects of the system:

1. Baseline Group (BG): Utilized the standard gpt-3.5-turbo model without any enhancements, serving as the control group for comparative analysis.

2. RAG-Enhanced Group (REG): Implemented the proposed RAG technology integration while maintaining the base model architecture.

3. Version Update Group (VUG): Employed the updated gpt-3.5-turbo-1106 model to assess the impact of model version improvements.

4. Fully Optimized Group (FOG): Combined both RAG technology and the latest model version, representing our complete proposed solution.

*4.3. Analysis of Model Comparison Results*

Table 6 summarizes the experimental results of different model configurations, comparing the accuracy and recall rates among baseline models (gpt-3.5-turbo and gpt-4.0-mini) and their RAG-enhanced variants, highlighting the significant performance improvements achieved through RAG technology integration.

**Table 6.** Model results

| Model | Accuracy rate | Recall |
|---|---|---|
| gpt-3.5-turbo | 55.6% | 78.3% |
| gpt-3.5-turbo+RAG | 63.7% | 88.7% |
| gpt-3.5-turbo-1106 | 59.3% | 85.4% |
| gpt-3.5-turbo-1106+RAG | 78.6% | 89.2% |
| gpt-4.0-mini | 68.2% | 86.9% |

The experimental results demonstrate significant variations in performance across different model configurations. In the baseline comparison, the gpt-3.5-turbo model achieved an accuracy rate of 55.6% and a recall rate of 78.3%, while the gpt-4.0-mini model showed improved performance with an accuracy rate of 68.2% and a recall rate of 86.9%. These results indicate that while basic models possess certain capabilities in handling financial data analysis tasks, their performance is constrained by limited understanding of domain-specific financial knowledge, suggesting substantial room for improvement.

The integration of RAG technology led to remarkable performance enhancements across all model variants. When comparing the two model groups - gpt-3.5-turbo with and without RAG, and gpt-3.5-turbo-1106 with and without RAG - we observed consistent improvements in both accuracy and recall metrics. Specifically, the gpt-3.5-turbo+RAG configuration achieved an accuracy rate of 63.7% and a recall rate of 88.7%, representing significant improvements over the baseline model. More notably, the gpt-3.5-turbo-1106+RAG configuration demonstrated exceptional performance, reaching an accuracy rate of 78.6% and a recall rate of 89.2%.

These improvements can be attributed to RAG technology's external knowledge retrieval mechanism, which enables models to access domain-specific financial knowledge efficiently. This enhanced knowledge access significantly improves the models' capability to understand and analyze financial data, effectively reducing errors stemming from knowledge limitations. The consistent improvement in both accuracy and recall rates across different model configurations underscores the crucial role of RAG technology in enhancing the performance of financial data analysis models.

## 5. Conclusion

In summary, this intelligent financial data analysis system based on LLM - RAG has demonstrated its effectiveness through a series of well - designed components and rigorous experiments. The data preprocessing module ensures the quality and consistency of the input data, laying a solid foundation for subsequent analysis. The vector storage and query retrieval mechanisms work in tandem to enable efficient access to and utilization of financial information, thereby enhancing the system's responsiveness to user queries.

The model comparison results clearly demonstrate the superiority of integrating RAG technology with large language models. The significant improvements in accuracy and recall rates following the incorporation of RAG indicate that it can effectively augment the models' knowledge base and enhancing their performance in processing financial data. This not only provides valuable insights for financial analysts and decision-makers but also holds considerable promise for advancing the development of intelligent financial services.

Future research could focus on further optimizing the integration of RAG andLLM, exploring more advanced retrieval algorithms and domain adaptation techniques. Additionally, expanding the dataset to encompass a broader range of financial markets and instruments could enhance the system's generalization ability. By continuously improving and innovating, we can expect this technology to play an increasingly important role in the financial industry, facilitating more informed and rational decision-making processes.

## References

[1] Arslan M, Munawar S, Cruz C. Business insights using RAG–LLMs: a review and case study[J]. Journal of Decision Systems, 2024: 1-30.

[2] Setty S, Thakkar H, Lee A, et al. Improving retrieval for rag based question answering models on financial documents[J]. arXiv preprint arXiv:2404.07221, 2024.

[3] Zhang B, Yang H, Zhou T, et al. Enhancing financial sentiment analysis via retrieval augmented large language models[C]//Proceedings of the fourth ACM international conference on AI in finance. 2023: 349-356.

[4] Wang S, Tan J, Dou Z, et al. OmniEval: An Omnidirectional and Automatic RAG Evaluation Benchmark in Financial Domain[J]. arXiv preprint arXiv:2412.13018, 2024.

[5] Lim J H, Suh J W. A Study on implementing the most optimized RAG system for financial document using AutoRAG[C]//Annual Conference of KIPS. Korea Information Processing Society, 2024: 521-522.

[6] Mao K, Liu Z, Qian H, et al. RAG-Studio: Towards In-Domain Adaptation of Retrieval Augmented Generation Through Self-Alignment[C]//Findings of the Association for Computational Linguistics: EMNLP 2024. 2024: 725-735.

[7] Yepes A J, You Y, Milczek J, et al. Financial report chunking for effective retrieval augmented generation[J]. arXiv preprint arXiv:2402.05131, 2024.